\documentclass[5p]{elsarticle}

\usepackage{graphicx}
\usepackage{epsfig}

\usepackage{amssymb}
\usepackage{bm, bbm}  

\newcommand{\bei}{\begin{itemize}}
\newcommand{\eei}{\end{itemize}}
\newcommand{\beq}{\begin{equation}}
\newcommand{\eeq}{\end{equation}}
\newcommand{\beqn}{\begin{eqnarray}}
\newcommand{\eeqn}{\end{eqnarray}}

\newcommand{\nn}{\nonumber}
\newcommand{\bq}{{\bf q}}
\newcommand{\bk}{{\bf k}}
\newcommand{\bK}{{\bf K}}
\newcommand{\bA}{{\bf A}}

\newcommand{\bR}{{\bf R}}

\newcommand{\br}{{\bf r}}
\newcommand{\be}{{\bf e}}
\newcommand{\ba}{{\bf a}}

\newcommand{\ua}{\uparrow}
\newcommand{\da}{\downarrow}

\newcommand{\ahat}{\hat{a}}
\newcommand{\bhat}{\hat{b}}

\newcommand{\etab}{\mbox{\boldmath $\eta $}}
\newcommand{\deltab}{\mbox{\boldmath $\delta $}}
\newcommand{\Pib}{\mbox{\boldmath $\Pi $}}

\usepackage{color}

\begin{document}

\begin{frontmatter}



\title{The quantum Hall effect in graphene -- a theoretical perspective}


\author[mark]{M. O. Goerbig}

\address[mark]{Laboratoire de Physique des Solides, Univ. Paris-Sud,
CNRS, UMR 8502, F-91405 Orsay Cedex, France}

\begin{abstract}

This short theoretical review deals with some essential ingredients for the understanding of the quantum Hall effect in graphene
in comparison with the effect in conventional two-dimensional electron systems with a parabolic band dispersion. The main difference 
between the two systems stems from the ``ultra-relativistic'' character of the low-energy carriers in graphene, which are described
in terms of a Dirac equation, as compared to the non-relativistic Schr\"odinger equation used for electrons with a parabolic band
dispersion. In spite of this fundamental difference, the Hall resistance quantisation is universal in the sense that it is given 
in terms of the universal constant $h/e^2$ and an integer number, regardless of whether the charge carriers are characterised by
Galilean or Lorentz invariance, for non-relativistic or relativistic carriers, respectively.

\end{abstract}

\end{frontmatter}

\section{Introduction}
\label{intro}

The integer quantum Hall effect (IQHE), discovered in 1980 \cite{KvK},
is a universal property of two-dimensional (2D) electrons in a strong magnetic field. At sufficiently
low temperatures, the effect manifests itself by a vanishing longitudinal resistance accompanied by a plateau in the transverse
(Hall) resistance with a value that is solely determined by the universal constant $h/e^2$ and an integer number. 
Whereas the main features of this effect may be understood in the framework of Landau quantisation of non-interacting 
2D electrons in a perpendicular magnetic field, its fractional counterpart, the fractional quantum Hall effect (FQHE)
discovered in 1982 \cite{TSG}, is induced by the mutual Coulomb interaction between the electrons. 

The experimental and theoretical investigation of the quantum Hall effect has experienced an unexpected revival with 
the discovery in 2005 of a particular IQHE in graphene (2D graphite) \cite{novoselov,zhang}. As compared to 2D electron
systems studied so far, the low-energy electronic properties in graphene are described by a relativistic Dirac equation 
for massless 2D particles rather than by the usual non-relativistic Schr\"odinger equation for lattice electrons with
a non-vanishing band mass. A direct consequence of this difference is an unconventional sequence of Hall plateaus, which 
are centred around the values of $\nu=\pm 2(2n+1)$ for the ratio $\nu=n_{el}/n_B$ between the electronic density $n_{el}$
and the density $n_B=eB/h$ of flux quanta threading the 2D surface. It is therefore natural to ask whether
both effects, the IQHE in conventional (non-relativistic) 2D electron gases and that in graphene, reveal the same 
universal properties, namely in view of the metrologically relevant Hall-resistance quantisation, or whether there are 
significant differences. 

The scope of the present short review is a comparison between the IQHE in graphene and that in conventional 2D electron systems.
We restrict ourselves to the discussion of non-interacting electrons, whereas the discussion of electronic interactions and of the 
recently discovered FQHE in graphene \cite{du,bolotin} would merit a review on its own. In Sec. \ref{sec1} we present the electronic
band structure of graphene and the relativistic Landau-level quantisation in the presence of a perpendicular magnetic field. 
The basic understanding of the IQHE in diffusive graphene samples is discussed in Sec. \ref{sec2}, in terms of (semi-classical)
electron localisation due to impurities and the particular form of the electronic confinement in graphene. Section \ref{sec3}
is devoted to the spin-valley degeneracy, namely in the zero-energy Landau level which is particular to graphene.

\section{Basics of graphene}
\label{sec1}

\begin{figure}
\centering
\includegraphics[width=6.5cm,angle=0]{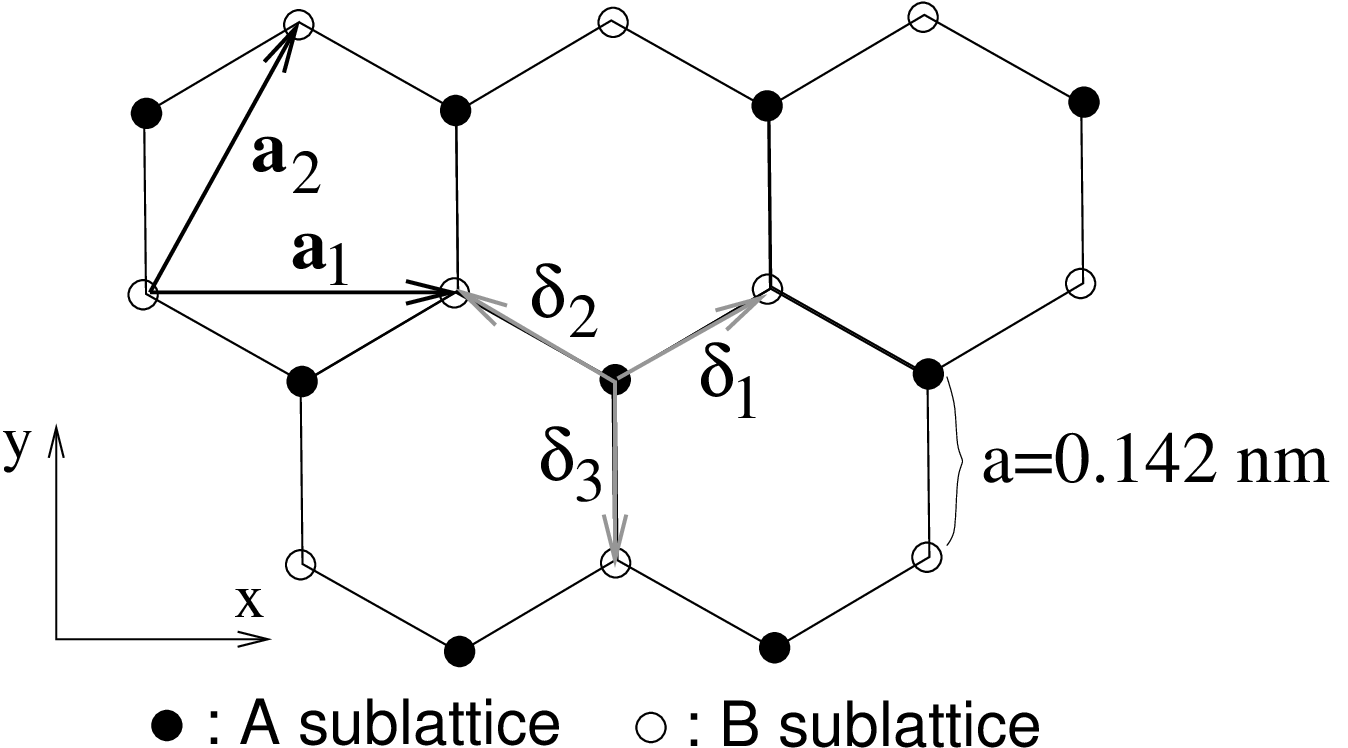}
\includegraphics[width=7.5cm,angle=0]{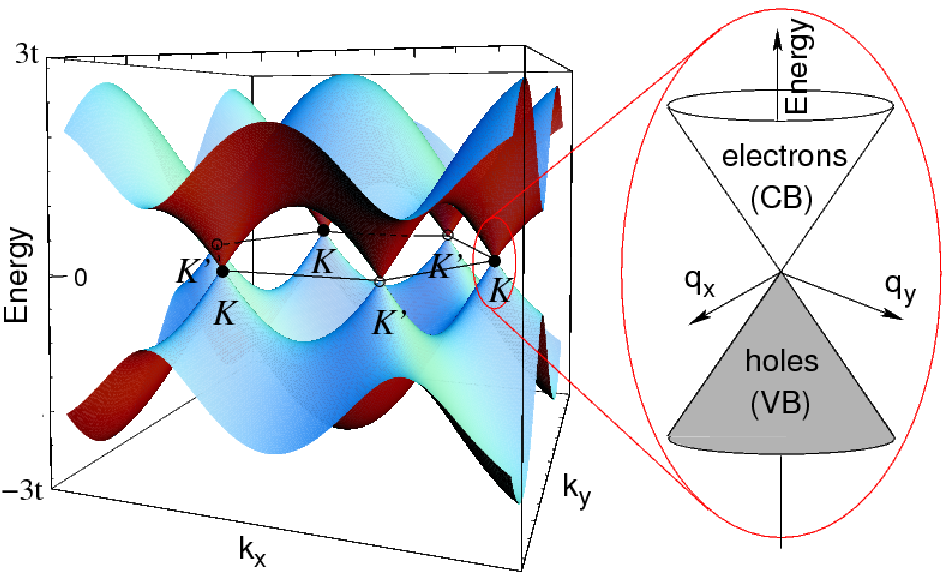}
\caption{\footnotesize{{\sl Upper panel:} honeycomb lattice. 
The vectors $\deltab_1$, $\deltab_2$,
and $\deltab_3$ connect {\sl nn} carbon atoms, separated by a distance $a=0.142$ nm.
The vectors $\ba_1$ and $\ba_2$ are basis vectors of the triangular Bravais
lattice. {\sl Lower panel:} electronic bands obtained from the tight-binding model with
nearest-neighbour hopping. The valence band touches the conduction band in the two inequivalent BZ corners $K$ and
$K'$. For undoped graphene, the Fermi energy lies precisely at the contact points, and the band dispersion in the vicinity of these
points is of conical shape.
}}
\label{fig01}
\end{figure}

From a crystallographic point of view, graphene is a 2D honeycomb lattice of carbon atoms that consists of two
triangular sublattices (Fig. \ref{fig01}, upper panel). 
Its basic electronic properties are readily understood within a simple tight-binding model 
in which one restricts the electron hopping to nearest-neighbour sites \cite{wallace}. The Hamiltonian resulting
from a decomposition into Bloch states with a lattice momentum $\bk$ reads
\beq\label{eq:Ham}
H_{\bk}=-t\left(\begin{array}{cc}
0 & \gamma_{\bk}^* \\ \gamma_{\bk} & 0
\end{array}\right),
\eeq
where $t\simeq 3$ eV is the hopping energy and the $2\times 2$ matrix reflects the 2 sublattices. The fact that
nearest-neighbour hopping involves sites on the two different sublattices makes the matrix off-diagonal, in terms
of the sum of phase factors $\gamma_{\bk}=\sum_{j=1}^3\exp(i\bk\cdot\deltab_j)$, where the vectors 
\beqn\label{eq1:02}
&&\deltab_1=\frac{a}{2}\left(\sqrt{3}\be_x+\be_y\right), \
\nn
\deltab_2=\frac{a}{2}\left(-\sqrt{3}\be_x+\be_y\right), \\
{\rm and}&&\deltab_3=-a\be_y,
\eeqn
connect a site on the A sublattice to its three nearest neighbours (see Fig. \ref{fig01}, upper panel). 

The dispersion relation, obtained from the diagonalisation of Hamiltonian (\ref{eq:Ham}), is depicted in the lower panel
of Fig. \ref{fig01}. One notices that the two bands touch each other in the corners of the first Brillouin zone (BZ),\footnote{
Although one notices six corners in the figure, only two of them are crystallographically inequivalent, i.e. not connected
by a reciprocal lattice vector.} which are labeled by $K$ and $K'$ and around which the dispersion relation is linear. In the
absence of doping, the Fermi level $E_F$ lies exactly in these contact points, but it may be varied either with the help of a 
backgate via the field-effect or by chemical doping. In order to describe the low-energy electronic
properties, which are typically restricted to the $100\, {\rm meV}\ll t$ regime, one may use a simplified linearised Hamiltonian
that is obtained from a series expansion of the factors 
\beq
\gamma_{\bq}^{\pm}\equiv \gamma_{\bk=\pm \bK+\bq} \simeq \mp \frac{3a}{2}(q_x \pm i q_y).
\eeq
Here $a=0.142$ nm is the distance between nearest-neigh\-bour carbon atoms and $\xi\bK=\xi (4\pi/3\sqrt{3}a)\be_x$ the
position of a $K$ ($\xi=+$) and a $K'$ point ($\xi=-$), respectively. The linearised Hamiltonian then reduces to a 2D Dirac Hamiltonian
for massless particles
\beq\label{eq2:32}
H_{\bq}^{{\rm eff},\xi}=\xi\hbar v_F(q_x\sigma_{AB}^x + \xi q_y\sigma_{AB}^y),
\eeq
where $\sigma_{AB}^x$ and $\sigma_{AB}^y$ are Pauli matrices associated with the sublattice ``spin'',
$v_F=3ta/2\hbar$ is the Fermi velocity, which is roughly 300 times smaller than the speed of light. 
Notice that the resulting dispersion relation 
$\epsilon_{\xi,\lambda\bq}=\lambda \hbar v_F|\bq|$, in terms of the band index $\lambda=\pm$ is independent of the valley index $\xi$, 
and one thus obtains a twofold valley degeneracy which adds to the twofold spin degeneracy.

In order to take into account the coupling to a perpendicular magnetic field $B$, one may use the so-called 
Peierls substitution, which is valid as long as the characteristic magnetic length $l_B=\sqrt{\hbar/eB}$ is much
larger than the lattice spacing. This is the case for experimentally accessible fields 
since $a/l_B\simeq 0.005\times \sqrt{B{\rm [T]}}$. The Peierls substitution consists of replacing 
the wave vector $\bq$ in the linearised Hamiltonian (\ref{eq2:32}) by the gauge-invariant kinetic momentum,\footnote{
We have chosen the electron charge to be $-e$ such that $e$ is the positive elementary charge.}
$\bq\rightarrow \Pib/\hbar\equiv -i\nabla + e\bA(\br)/\hbar$, where $\bA(\br)$ is the vector potential that
generates the magnetic field $B\be_z=\nabla\times\bA(\br)$. Canonical quantisation, with the commutation relation
$[x_{\mu},p_{\nu}]=i\hbar\delta_{\mu,\nu}$ between the components $x_{\mu}$ of the position operator and those
$p_{\mu}=-i\hbar\partial/\partial x_{\mu}$ of the canonical momentum operator ($\mu,\nu=x,y$ for the 2D plane),
yields the non-commutativity between the components of the kinetic momentum
\beq
[\Pi_x,\Pi_y]=-i\frac{\hbar^2}{l_B^2},
\eeq
such that these components may be viewed as conjugate. One may therefore introduce the convenient ladder operators
\beq\label{ladder}
\ahat = \frac{l_B}{\sqrt{2}\hbar}\left(\Pi_x - i\Pi_y\right) ~ {\rm and} ~
\ahat^{\dagger} = \frac{l_B}{\sqrt{2}\hbar}\left(\Pi_x + i\Pi_y\right),
\eeq
which satisfy the usual commutation relation $[\ahat,\ahat^{\dagger}]=1$, as in the case of the harmonic oscillator.
In terms of these ladder operators, the linearised Hamiltonian in a magnetic field becomes
\beq\label{HamLadD}
H_B^{\xi}= 
\xi\sqrt{2}\frac{\hbar v_F}{l_B}\left(\begin{array}{cc}
0 & \ahat \\ \ahat^{\dagger} & 0
        \end{array}\right) ,
\eeq
where, as compared to Hamiltonian (\ref{eq2:32}), we have interchanged the A and B component in the spinors describing
electrons around the $K'$ point ($\xi=-$).

\subsection{Relativistic Landau levels}

The solution of the eigenvalue equation $H_{B}^{\xi}\psi_n=\epsilon_{\lambda n}\psi_n$,
in terms of the 2-spinors 
\beq
\psi_n=\left(\begin{array}{c} u_n \\ v_n \end{array} \right).
\eeq
yields the level spectrum of graphene electrons in a magnetic field, 
\beq\label{eigen2}
\ahat^{\dagger}\ahat\, v_n = \left(\frac{\epsilon_{\lambda n}}{\sqrt{2}\hbar v_F/l_B}\right)^2 v_n,
\eeq
which indicates that $v_n\propto |n\rangle$ is an eigenstate of the number operator $\ahat^{\dagger}\ahat$, 
$\ahat^{\dagger}\ahat|n\rangle=n|n\rangle$. One thus obtains
the {\sl relativistic Landau levels}
\beq\label{RelLLs}
\epsilon_{\lambda n} = \lambda \frac{\hbar v_F}{l_B}\sqrt{2n},
\eeq
where $\lambda=\pm$ denotes the levels with positive and negative energy, respectively \cite{ML}. Furthermore,
the substitution of this result in the eigenvalue equation yields $u_n\propto \lambda \ahat |n\rangle$.
The term {\sl relativistic} is used to distinguish clearly the $\lambda\sqrt{Bn}$ dispersion of the levels,
obtained from the 2D Dirac equation for massless particles, from the conventional (non-relativistic) Landau levels,
which disperse linearly in $Bn$.

Another notable difference with respect to non-relativistic Landau levels in metals with parabolic bands is the presence
of a zero-energy Landau level with $n=0$. This level needs to be treated separately, and indeed the solution of the 
eigenvalue equation yields an eigenvector 
\beq\label{spinN0}
\psi_{\xi,n=0} = \left(\begin{array}{c} 0 \\ |n=0\rangle  \end{array}\right),
\eeq
with a single non-vanishing component. As a consequence, zero-energy states at the $K$ point are restricted to the B sublattice,
whereas those at the $K'$ have a non-vanishing weight only on the A sublattice. In Landau levels with $n\neq 0$, the eigenstates
\beq\label{spinN}
\psi_{\lambda,n\neq 0}^{\xi} = \frac{1}{\sqrt{2}}\left(\begin{array}{c} |n-1\rangle \\ \xi\lambda |n\rangle  \end{array}\right)
\eeq
are spinors in which both sublattices are equally populated, but the components correspond do different non-relativistic 
Landau states. 

\subsection{Degeneracy of Landau levels}

As in the non-relativistic case, the relativistic Landau levels in graphene are highly degenerate. This is 
a consequence of the existence of a second pair of conjugate variables, in addition to $\Pi_x$ and $\Pi_y$, which may be obtained
formally from a similar combination as for $\Pib$, but with a different relative sign, $\tilde{\Pib}\equiv -i\hbar \nabla - e\bA(\br)$.
This quantity has the dimension of a momentum but is, in contrast to $\Pib$, gauge-dependent. In spite of its formal character,
one may show that this quantity commutes, when choosing the symmetric gauge $\bA(\br)=B(-y,x,0)/2$, with the components 
of $\Pib$ and thus with the Hamiltonian (\ref{HamLadD}). The operator $\tilde{\Pib}$ represents thus a constant of motion.
However, its components do not commute among themselves,
$[\tilde{\Pi}_x,\tilde{\Pi}_y]=i\hbar^2/l_B^2$, such that one may introduce the ladder operators 
\beq\label{ladderGC}
\bhat = \frac{l_B}{\sqrt{2}\hbar}\left(\tilde{\Pi}_x + i\tilde{\Pi}_y\right) ~ {\rm and} ~
\bhat^{\dagger} = \frac{l_B}{\sqrt{2}\hbar}\left(\tilde{\Pi}_x - i\tilde{\Pi}_y\right),
\eeq
which generate a second quantum number $m$, with $\bhat^{\dagger}\bhat|m\rangle=m|m\rangle$. Moreover, 
the operator $\tilde{\Pib}$ is associated with the guiding centre, $\bR=\be_z\times\tilde{\Pib}/eB$,
which describes the centre of the cyclotron motion within a semi-classical picture. This property is very intuitive
because the centre of the cyclotron motion is indeed a constant of motion and thus commutes with the Hamiltonian. Furthermore,
because of the non-commutativity of the components of $\tilde{\Pib}$, the commutation relation for the guiding-centre
components reads 
\beq\label{eq:commGC}
[X,Y]=il_B^2, 
\eeq
which yields a non-commutative 2D geometry as well as the uncertainty relation
$\Delta X\Delta Y=2\pi l_B^2$. This is a remarkable result because it indicates that each quantum state is smeared over
a minimal surface $2\pi l_B^2=h/eB$ which happens to be the inverse of the flux density $n_B=eB/h$ measured in units
of the flux quantum $h/e$. The flux density therefore determines the degeneracy of each Landau level.

Notice that the above argument does not require a particular form of the level spectrum -- it is valid both for non-relativistic
and for relativistic Landau levels. The degeneracy of the Landau levels, apart from internal degrees of freedom such as the spin or
the fourfold spin-valley degeneracy in graphene, is therefore determined by the flux density, and one may introduce in both
cases the same filling factor, in terms of the electronic density $n_{el}$,
\beq\label{filling}
\nu=\frac{n_{el}}{n_B}=\frac{h n_{el}}{eB},
\eeq
which characterises the filling of the Landau levels.

\subsection{Confinement}

Until now, we have described the graphene sheet as an infinite 2D plane with no boundaries. However, in a realistic sample, which
is used in transport measurements, the electrons are confined to a delimited area. In graphene, the confinement depends on the
precise form of the edges and happens to be quite different from that of a conventional 2D electron gas \cite{BF06}. Indeed,
a simple electrostatic potential barrier is insufficient to confine electrons inside the graphene sheet, as a consequence of their
chiral properties and the resulting Klein tunneling that allows electrons to enter the gated region \cite{KleinT}. The main
features of electronic confinement in graphene may be understood with the help of the term
\beq\label{MassTerm}
V_{\rm conf}(y) = V(y)\, \sigma_{AB}^z=
\left(\begin{array}{cc} V(y) & 0 \\ 0 & - V(y) \end{array}\right),
\eeq
where we have chosen a confinement in the $y$-direction with a potential that vanishes in the interval $y_{min}<y<y_{max}$ and
diverges for $y\ll y_{min}$ and $y\gg y_{max}$. Formally the confinement potential has the form of a $y$-dependent mass term, i.e. 
the electrons acquire a mass $V(y)$ beyond the sample edges. Notice that Eq. (\ref{MassTerm}) represents a simplified form 
of electronic confinement in graphene that does not take into account the lattice structure at the edge (armchair or zigzag), 
which gives rise to 
a fine structure of the level spectrum there \cite{BF06}. Indeed, the valley degeneracy for $n\neq 0$ LLs
is lifted at an armchair edge such that 
the edge channels corresponding to the two different valleys are spatially separated, on the scale of $l_B$. However, the chirality
of the edge states remains unchanged such that this fine structure does not affect the quantisation of the Hall resistance. The latter
may therefore be described in the above simplified scheme of Dirac-fermion confinement (\ref{MassTerm}),
which corresponds to the zigzag edge.

\begin{figure}
\centering
\includegraphics[width=6.5cm,angle=0]{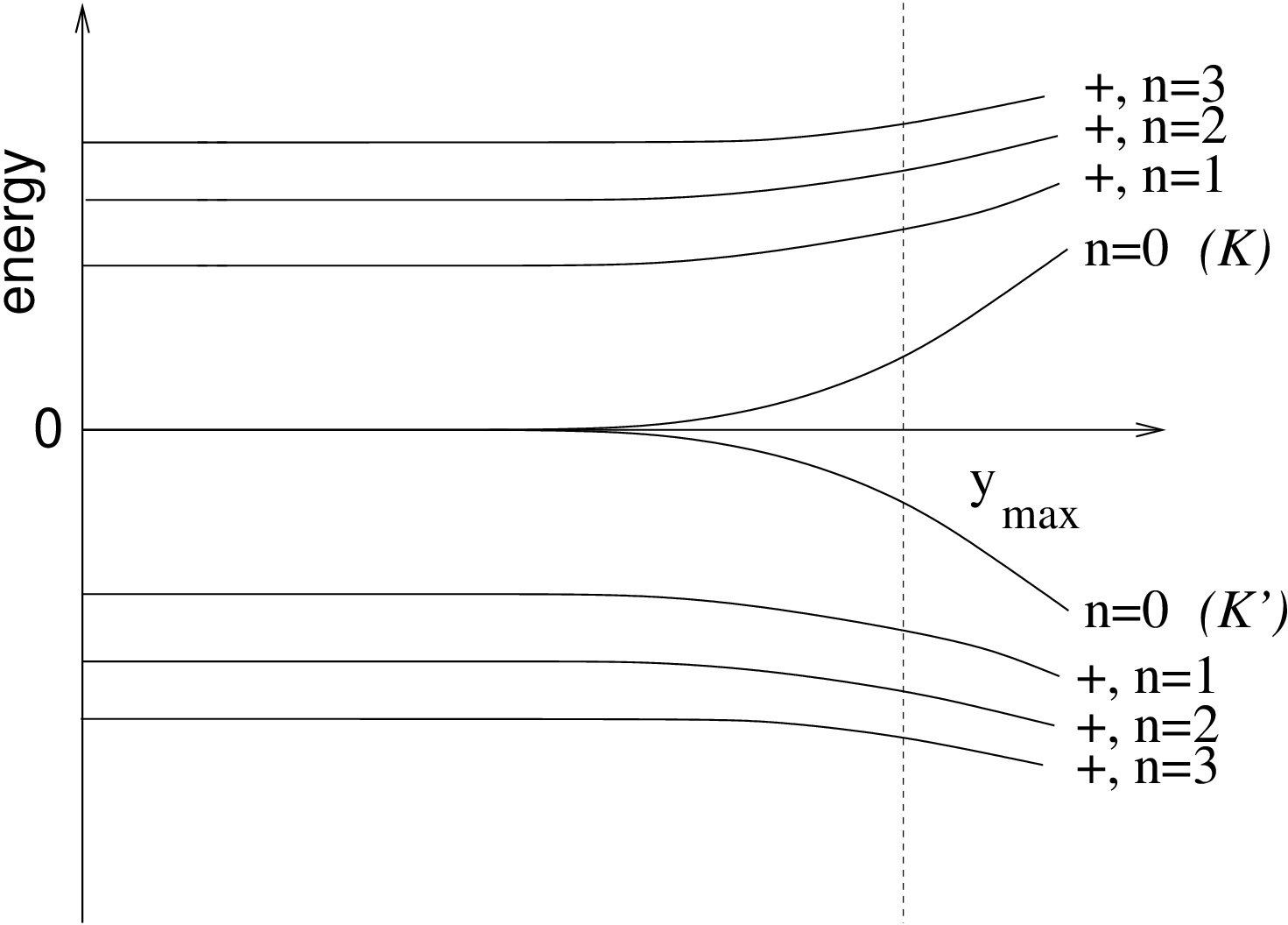}
\caption{\footnotesize{ 
Generic form of the confinement for relativistic Landau levels. 
Whereas electron-like Landau levels ($\lambda=+$) are bent upwards when approaching
the edge $y_{max}$, the hole-like levels ($\lambda=-$) are bent downwards. The behaviour of the zero-energy level ($n=0$)
depends on the valley index. In the $K$ valley it increases in energy, whereas it decreases in the other one ($K'$).
}}
\label{fig02}
\end{figure}

The particular choice of the confinement potential, which respects the translation invariance in the $x$-direction, allows one 
to solve the Hamiltonian $H_B^{\xi}+V_{\rm conf}(y)$
in the Landau gauge $\bA_L=B(-y,0,0)$, in which case the wave vector $k$ in the $x$-direction is a good
quantum number, and the quantum state is described essentially by a Gaussian that is centred around the coordinate $y_0=kl_B^2$.
One may therefore approximate the confinement potential by replacing the $y$-coordinate by its average value $y_0$,
$V(y)\simeq V(y_0=kl_B^2)$. Notice that the wave vector $k$ plays the same role as the guiding-centre quantum number $m$ 
in the symmetric gauge, and the $k$-dependence of the Hamiltonian thus lifts the Landau-level degeneracy, as may be seen
from the energy spectrum (\ref{RelLLs}) which becomes, in the presence of the term (\ref{MassTerm}),
\beq\label{MassSpecA}
\epsilon_{\lambda n,y_0;\xi} = \lambda \sqrt{V^2(y_0) + 2\frac{\hbar^2 v_F^2}{l_B^2} n}, 
\eeq
for $n\neq 0$ and both valleys $\xi=\pm$, whereas the $n=0$ LL is found to depend explicitly on the valley index $\xi$,
\beq\label{MassSpec0}
\epsilon_{n=0,y_0;\xi} = \xi V(y_0).
\eeq
The level spectrum is depicted in Fig. \ref{fig02}.

In order to summarise the differences and similarities between conventional 2D electron systems with a parabolic band and 
relativistic electrons with zero band mass in graphene and in order to prepare the discussion of the relativistic quantum Hall 
Hall effect, we notice that:
\begin{itemize}
\item[1.] The Landau level spectra are different. Apart from a different $B$-field dependence, graphene is characterised by
a zero-energy level ($n=0$) that is somewhat in between the electron-like levels in the conduction band ($n\neq 0$ with $\lambda=+$)
and the hole-like ones in the valence band (with $\lambda=-$);

\item[2.] The level degeneracy, due to the guiding-centre quantum number, is the same in both cases, $n_B$, and the filling is therefore
characterised by the same filling factor $\nu$. Notice, however, that $\nu=0$ in graphene describes the particle-hole-symmetric
charge-neutrality point, where the zero-energy Landau level $n=0$ is necessarily half-filled with electrons (or holes).

\item[3.] The internal spin-valley symmetry yields an additional fourfold degeneracy of each Landau level in graphene in contrast
to the twofold spin degeneracy in a conventional 2D electron gas.\footnote{The spin degeneracy is naturally lifted by the Zeeman effect,
which happens though to be characterised by an energy scale that is much smaller than $\hbar v_F/l_B$ or the typical interaction
energy $e^2/\epsilon l_B$.}

\item[4.] Confinement needs to be treated differently in graphene as compared to the conventional 2D electron gas, where it 
can be modeled by an electrostatic potential. In the case of graphene, one needs to use a mass confinement (\ref{MassTerm}) 
to restrict the Dirac electrons into a particular region.

\end{itemize}

\section{The quantum Hall effect}
\label{sec2}

In this section, we discuss the IQHE in diffusive conductors which turn out to be most relevant for metrological
issues since they provide large Hall plateaus
\cite{poirier}. In diffusive samples, the impurities break the translational invariance without which 
the Hall resistance would not be quantised because of a possible Lorentz transformation that yields the classical value of the 
Hall resistance \cite{girvin}.\footnote{Notice however that the quantum Hall effect also occurs in ballistic nanoscale samples, 
where the necessary translation-symmetry breaking is achieved by the sample boundaries.}
In order to understand the universality of the IQHE, as in conventional 2D electron systems, one requires the following
ingredients:
\bei
\item[1.] At certain (integer) filling factors, the ground state, which consists here of completely filled Landau levels, is separated
by a gap from its charged excitations. According to the discussion in the previous section, 
the gap is simply the energy difference between the last
occupied level and the next unoccupied one. In the case of graphene, 
the IQHE condition is then fulfilled when $\nu=\pm2(2n+1)=\pm 2,\pm 6, \pm 10, ...$. The steps in units of 4 reflects precisely
the fourfold spin-valley degeneracy in graphene, whereas the ``offset'' of 2 is due to the fact that at $\nu=0$ the zero-energy 
level is only half-filled, and the highly degenerate ground state therefore does not have a gap to charged excitations.\footnote{
This statement is only valid if one considers the kinetic energy. Interactions or external fields, such as the Zeeman effect,
are capable of lifting the ground-state degeneracy.}

\item[2.] Each completely filled spin-valley subbranch of a 
Landau level may be viewed as a current-carrying mode with a perfect transmission that contributes
one quantum of conductance $e^2/h$ to the charge transport. 

\item[3.] When changing the Landau-level occupation, either by changing the electronic density or the magnetic field which modifies the
number of available states per level, the additional electrons are forced to occupy the next higher Landau level (or additional holes
in the last occupied one) and are localised (semi-classically) by the sample impurities. These additional charges therefore do not
contribute to the charge transport, such that the transport characteristics, i.e. the resistances, remain unchanged. We are
thus confronted with the unusual situation that impurities that normally blur the precise quantisation 
of the energy levels are essential for the 
occurence of the IQHE, but the effect remains universal in the sense that the precise impurity distribution is not relevant 
for this quantum effect.

\eei

The first point has already been discussed in the previous section. In order to illustrate the second point, we calculate
the current of a single completely filled Landau level, 
which flows between two contacts (source and drain),\footnote{Other contacts may be present but are considered as
contacts with an infinite internal resistance such that no electrons leak in or out. For a more complete discussion of
multi-terminal configurations, see B\"uttiker's review \cite{Butt}.} 
\beq
I_n=-\frac{g e}{\hbar L}\sum_k\frac{\partial \epsilon_{\lambda n,y_0=kl_B^2;\xi}}{\partial k},
\eeq
which may be evaluated with the help of periodic boundary conditions in the $x$-direction, $k=2\pi m/L$, such that
$\partial \epsilon_{\lambda n,y_0=kl_B^2;\xi}/\partial k=(L/2\pi \hbar)\Delta\epsilon_{\lambda n,y_0;\xi}/\Delta m$. 
Here, $g=4$ represents the fourfold spin valley degeneracy of a graphene Landau level. One
notices then that the different contributions in the sum are canceled apart from the boundary terms at $y_{min}$ and 
$y_{max}$, such that the current reads
\beq
I_n=-\frac{g e}{h}(\mu_{max}-\mu_{min}),
\eeq
where $\mu_{max/min}=\epsilon_{\lambda n,y_{max/min}l_B^2;\xi}$
are the chemical potentials $\mu_{max/min}$ at the edges, the difference of which is simply the voltage $-eV=\mu_{max}-\mu_{min}$
between the edges.  This yields $I_n=(g e^2/h)V$, 
i.e. each graphene Landau level above the charge neutrality point ($n\neq 0$) contributes 
\beq\label{eq:condN}
G_{n\neq 0}=g\frac{e^2}{h}
\eeq
to the (two-terminal) 
conductance, as stated above. In the case of the zero-energy level $n=0$, only half of it is electron-like, such that the level
contributes only half of the value (\ref{eq:condN}), i.e. $G_0=ge^2/2h$, to the conductance. 

The third essential ingredient mentioned above is concerned with the (semi-classical) localisation properties 
in the presence of a disorder potential. Quite generally, such a potential may be represented
\beq\label{eq:imp}
V_{imp}(\br)=\sum_{\mu,\nu,\sigma}V_{\mu,\nu,\sigma}(\br)\sigma_{AB}^{\mu}\otimes\tau_{V}^{\nu}\otimes\tau_{spin}^{\sigma},
\eeq
in terms of three ``spin'' operators, the sublattice $\sigma_{AB}^{\mu}$, the valley $\tau_{V}^{\nu}$ and the physical
spin $\tau_{spin}^{\sigma}$, where $\mu,\nu,\sigma=0,x,y,z$.
The most familiar component, apart from the long-range electrostatic potential associated with the component 
$\sigma_{AB}^0\otimes\tau_{V}^0\otimes\tau_{spin}^0$, is probably the true spin component 
$\sigma_{AB}^0\otimes\tau_{V}^0\otimes\tau_{spin}^{\sigma}$, which may be generated e.g. by magnetic impurities. The 
other components are proper to graphene. Indeed, the valley component $\sigma_{AB}^0\otimes\tau_{V}^{\nu}\otimes\tau_{spin}^0$
would be generated by short-range scatterers that couple the different valleys and that have non-negligible Fourier
components at a wave vector $\sim 1/a$, whereas the component $\sigma_{AB}^{\mu}\otimes\tau_{V}^0\otimes\tau_{spin}^0$ with
$\mu=x,y$ corresponds to gauge-field fluctuations.\footnote{As we have already seen in the previous section, $\mu=z$ would
correspond to a local mass term.}
The latter may be generated by bond disorder, such as ripples \cite{meyer},
which couple to the electronic degrees of freedom via a modification of the hopping parameter, $(\partial t/\partial a)\delta a$,
where $\delta a$ represents a deformation of a carbon-carbon bond. 

Similarly to the above-mentioned confinement, the disorder potential $V_{imp}(\br)$ lifts the Landau-level degeneracy,
and its effect may be studied from the Heisenberg equations of motion for the guiding-centre operator, which are obtained
from a series expansion of the potential in the cyclotron coordinate
$\etab=\br-\bR$. The leading-order term reads $V_{imp}(\br)\simeq V_{imp}(\bR)$.\footnote{Higher-order
terms, which become important in the case of rapidly varying potential $|\nabla V_{imp}|\sim \Delta_{n}/l_B$ in terms
of the level spacing $\Delta_n\simeq \hbar (v_F/l_B)/\sqrt{2n}$, may be systematically taken into account in a vortex basis
\cite{champel}.}
With the help of the commutation relation (\ref{eq:commGC}), one obtains 
\beq\label{eq:semicl}
\dot{\bR}=\frac{1}{eB}\sum_{\mu,\nu,\sigma}\left[\nabla_{\bR}V_{\mu,\nu,\sigma}\times \be_z\right]
\sigma_{AB}^{\mu}\otimes\tau_{V}^{\nu}\otimes\tau_{spin}^{\sigma},
\eeq
where $\nabla_{\bR}$ denotes the gradient with respect to the guiding centre. Similarly to the 2D electron gas
with a parabolic band in semi-conductor heterostructures, the Heisenberg equations of motion induce a semi-classical
Hall drift of the guiding centre, $\langle \dot{\bR}\rangle \perp \nabla_{\bR}V$, even if the physical properties are richer in
graphene due to the internal degrees of freedom. Notice that inter-Landau-level excitations are not treated on this level
of the approximation, where we have substituted $\br\rightarrow \bR$ in the argument of the impurity potential, but
they occur when higher-order corrections in $\etab$ are taken into account. 

\begin{figure}
\centering
\includegraphics[width=8.5cm,angle=0]{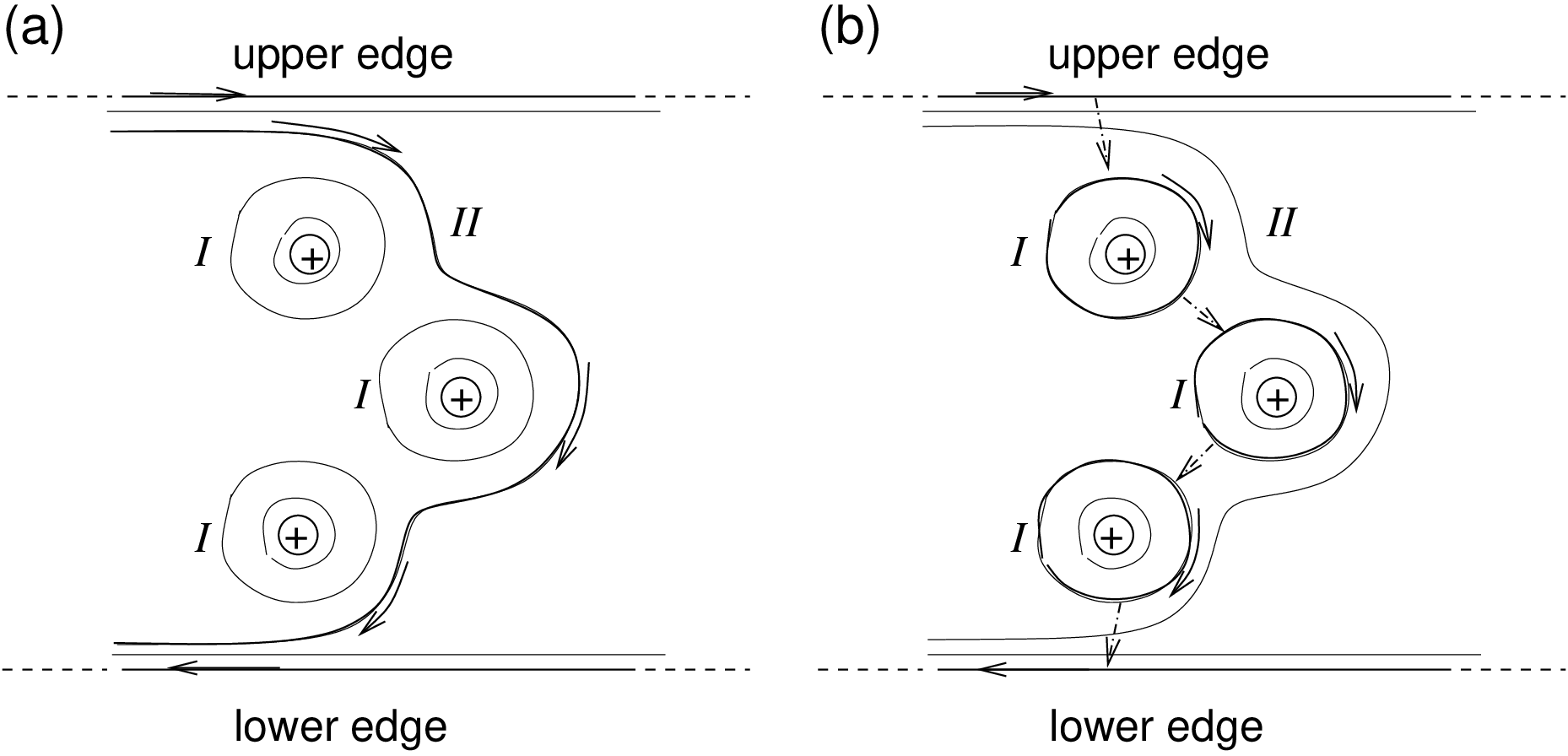}
\caption{\footnotesize{ {\sl (a)} Classical percolation. An electron propagating from left to right at the upper edge can
reach the lower edge and thus be back-scattered via a quantum state represented by an open equipotential line (II), whereas
the closed lines (I) do not contribute to the current transport.  
{\sl (b)} Enhanced percolation via quantum tunneling. Close to the classical percolation threshold, the electron puddles are
sufficiently close such that an electron may tunnel from one puddle border to the other. An electron at the upper edge 
can thus reach the lower edge via multiple tunneling events (variable-range hopping).
}}
\label{fig03}
\end{figure}

As a consequence of the semi-classical Hall drift, the electronic motion is bound to equipotential lines of the disorder
potential. The electrons occupying quantum states corresponding to closed lines may be viewed as localised states that 
do not contribute to the electronic transport, whereas those occupying open lines are extended states that can carry current
between spatially separated electronic contacts. A particular form of these extended states are the so-called edge states 
which are formed at the sample edges as a consequence of the above-mentioned confinement potential and which
turn out to be the relevant current-carrying states in the vicinity of the magic filling factors $\nu=\pm2(2n+1)$. 
As one may see from the semi-classical equations of motion for the guiding centre, the magnetic field and the gradient
of the confinement potential impose a well-defined direction of the electronic motion, and the edge states are therefore
chiral. At filling factors that slightly mismatch these magic values, additional electrons (or holes) first occupy
the lowest-energy states of the adjacent Landau level, which may be viewed as small closed orbits encircling the bottom
of the valleys of the potential landscape. The additional charges are thus localised and the
transport characteristics (the measured resistances) do not change when varying slightly the filling factor.
Therefore, one obtains a plateau in the Hall resistance at the same quantised value $R_H=\pm (h/e^2)\times 1/2(2n+1)$
as for the situation where the filling factor is precisely $\nu=\pm 2(2n+1)$ \cite{peres06}, in agreement with 
the experimental observation of the IQHE in graphene \cite{novoselov,zhang}.

The situation changes when one varies the filling factor more importantly such that a Landau level is roughly half-filled,
in which case the few extended states in the bulk represented by open equipotential lines (see Fig. \ref{fig03}) are occupied
and separating a puddle filled with electrons from an empty region. If these lines connect opposite sample edges, the associated
quantum states serve as short-circuits by which an electron from one sample edge can leak to the other one. In this 
case, the chemical potential is no longer constant along the sample edge between the source and the drain, and additional
contacts in between could be used to measure the voltage drop and thus a non-zero longitudinal resistance. At the same moment,
the Hall resistance measured between contacts on opposite sample edges is no longer quantised. 

This classical percolation picture of the plateau transition needs to be modified when quantum mechanical tunneling
is taken into account. Already below the classical percolation threshold, where extended equipotential lines start
to be occupied, the electron (or hole) puddles increase in size and are separated only by small unoccupied bottlenecks. 
Electrons may therefore quantum-mechanically tunnel from one puddle boundary to an adjacent one, and an electron that is 
originally propagating along one sample edge may reach the opposite sample edge via multiple tunneling events through the bulk. 
This picture is at the origin of the variable-range-hopping model \cite{VRH} that has recently been tested sucessfully
in graphene \cite{keyan}.

To summarise, the IQHE in graphene has a clear relativistic fingerprint due to the particular succession of
Hall plateaus at the filling factors $\nu=\pm 2(2n+1)$, as a consequence of the relativistic Landau-level quantisation. However, 
it shows the same universality as the IQHE in conventional 2D electron gases with a parabolic band dispersion -- the quantisation
of the Hall resistance is determined by a universal constant $h/e^2$ and integer numbers. 
The universality and the high metrological precision of the Hall resistance quantisation has recently been confirmed 
experimentally \cite{tza} in large epitaxial graphene flakes \cite{deheer}.
Furthermore, the impurity potential
plays the same essential role in a diffusive graphene sample as in a conventional 2D electron gas: it is necessary to pin
the resistances to their universal values via the above-mentioned semi-classical localisation, whereas its precise form and 
the microscopic distribution of the scatterers is of no importance for the quantisation. 

This universality between the 
relativistic and the non-relativistic variant of the IQHE may furthermore be understood from the electron dynamics once restricted
to a single Landau level. Whereas the electrons, in the absence of a magnetic field, are originally governed by different types
of space-time invariance (Galilean invariance in the case of a non-relativistic 2D electron gas, in contrast to Lorentz invariance
for electrons in graphene), the translation symmetry for electrons within a single Landau level respects neither of these invariances;
it is rather determined by the magnetic
translations which are induced by the commutation relations (\ref{eq:commGC}) for the guiding-centre coordinates, both in the case of
relativistic and non-relativistic electrons. The Heisenberg equations of motion (\ref{eq:semicl}) are, apart from the more complicated 
internal structure in the case of graphene,\footnote{The different possible aspects of the internal structure of the scattering
potential remains though to be understood in more detail in the presence of a magnetic field.}
indeed of the same form as in conventional 2D electron gases, and one may therefore expect that not only the Hall quantisation,
but also the nature of the Hall transitions and the associated critical exponents are universal.

\section{Spin-valley degeneracy lifting}
\label{sec3}

Additional Hall plateaus may arise when the fourfold spin-valley degeneracy is lifted. If the Coulomb interaction is taken into account,
this degeneracy is lifted due to the spontaneous formation of a generalised quantum Hall ferromagnetic state in both the spin and
the valley channel, at integer filling factors different from $\nu=\pm 2(2n+1)$ \cite{goerbigRev}. However, the approximate SU(4) 
spin-valley symmetry of the Coulomb interaction does not allow one to discriminate whether the spin or the valley degeneracy 
is preferentially lifted. The hierarchy of the degeneracy lifting, also in the presence of interactions, is therefore determined
by external symmetry-breaking fields even if these turn out to be much smaller in energy than the typical interaction-energy scale
$e^2/\epsilon l_B$.

The simplest and most familiar of these symmetry-breaking fields is definitely the Zeeman effect that is described by the 
term 
\beq
H_{Z}=\Delta_Z \sigma_{AB}^0 \otimes \tau_V^{0}\otimes \tau_{spin}^z,
\eeq
where $\Delta_Z\simeq 0.1\times B{\rm [T]}$ meV is the typical energy scale for a $g$-factor that has been determined as
$\sim 2$ \cite{zhang2}. The Zeeman effect separates each Landau level into two spin branches, and additional Hall plateaus may
then occur even in the absence of Coulomb interactions. Similarly, one may postulate a ``valley'' Zeeman effect that couples
to the valley pseudospin. However, such a valley Zeeman effect is more involved than the natural Zeeman effect, which 
couples to the physical spin. Recently, it has been argued that a spontaneous Kekul\'e-type deformation of the honeycomb lattice 
might occur as a consequence of the coupling between the electronic degrees of freedom and inplane optical phonons \cite{Hou09}.
The deformation may be viewed as consisting of deformed hexagons, in which every second bond is shortened with respect to the others.
It enlarges the unit cell by a factor of three such that a reciprocal lattice vector of the new underlying Bravais lattice is commensurate
with the wave vector connecting the $K$ and $K'$ points. As a consequence, the Kekul\'e-type deformation couples
to the valley degree of freedom and yields a term of the form \cite{Hou09}
\beq\label{eq:kek}
H_K=\sum_{\nu=x,y}\Delta^{\nu}\sigma_{AB}^0\otimes \tau_V^{\nu}\otimes \tau_{spin}^0,
\eeq
where $\Delta^{\nu}$ are real parameters and represent energies that are on the same order of magnitude as (though slightly
larger than) the typical Zeeman energy in graphene. In contrast to the latter, the contribution (\ref{eq:kek}) does not lift
the valley degeneracy in all relativistic Landau levels, but only in the zero-energy level ($n=0$) -- indeed a diagonalisation
of the Hamiltonian $H_B^{\xi}+H_K$ yields the same energy spectrum as the massive Dirac Hamiltonian given by Eqs. (\ref{MassSpecA})
and (\ref{MassSpec0}) if one replaces $V(y_0)$ by $\Delta_K=\sqrt{\Delta^{x2}+\Delta^{y2}}$. The Kekul\'e-type deformation therefore 
yields the same degeneracy lifting as the previously discussed out-of-plane deformation which gives rise to a mass gap in the presence
of a magnetic field \cite{FL}, and we use $\Delta_K$ to describe a general valley-degeneracy lifting in the remainder of this 
review.

\begin{figure}
\centering
\includegraphics[width=8.8cm,angle=0]{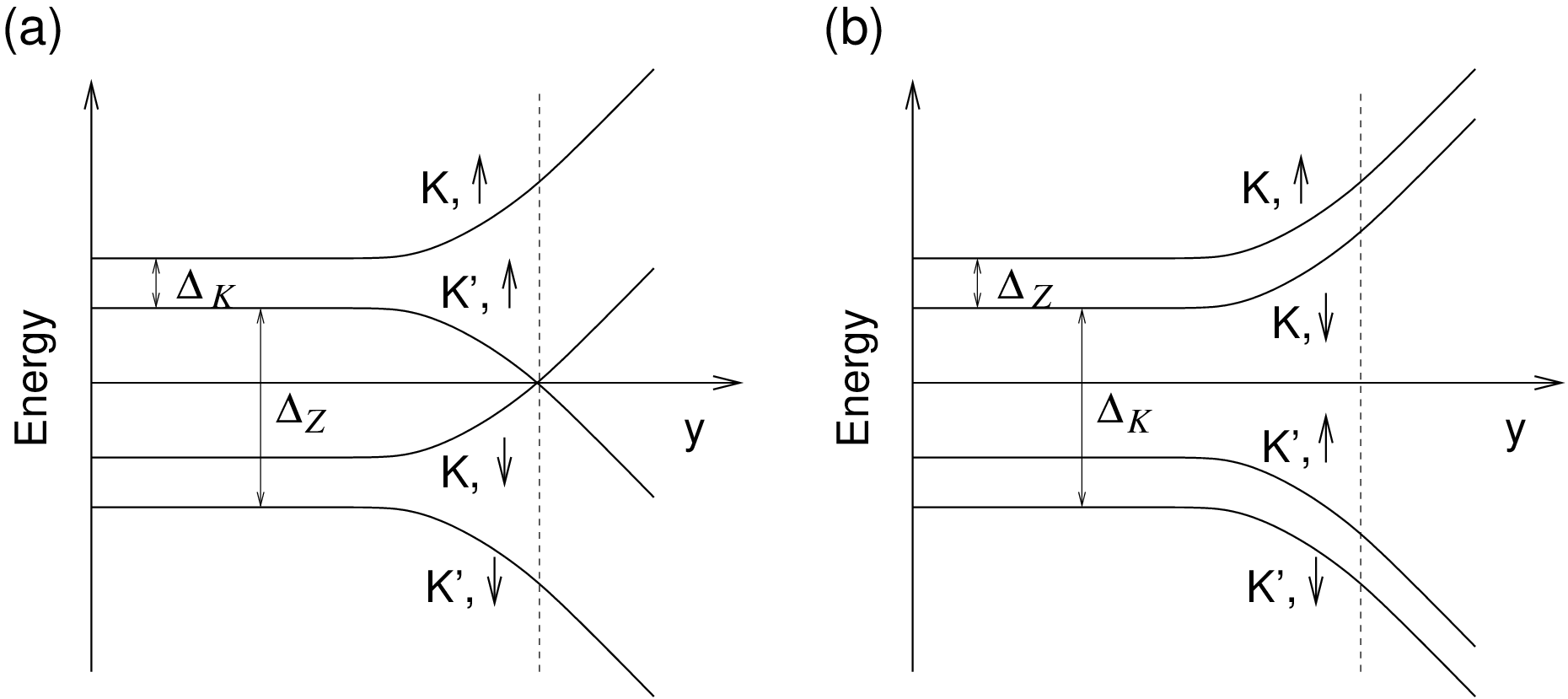}
\caption{\footnotesize{Scenarios for the lifted spin-valley degeneracy at $\nu=0$.
{\sl (a)} $\Delta_Z>\Delta_K$ in the bulk. When approaching the edge, the energy difference between the two valleys 
increases drastically, and two levels ($K',\ua$) and ($K,\da$) cross the Fermi energy at the edge depicted by the dashed line
(Quantum Hall state).
{\sl (b)} $\Delta_K>\Delta_Z$ in the bulk. 
The $K$ subbranches are already located above the Fermi energy, and those 
of $K'$ below, such that the energy difference is simply increased when approaching the edge with no states crossing the
Fermi energy (Insulator).
}} 
\label{fig04}
\end{figure}

This discussion indicates that the zero-energy level needs to be treated differently from the other Landau levels in graphene.
The hierarchy of the energy scales $\Delta_Z$ and $\Delta_K$ do not only determine the succession of the spin-valley 
degeneracy lifting in $n=0$ but also the conduction properties, namely at the charge neutrality point $\nu=0$ (see Fig. \ref{fig04}).
Indeed, if the Zeeman energy is larger than the one associated with the valley coupling, both valley subbranches ($K,\sigma=\ua$) and 
($K',\ua$) of the spin branch of $n=0$ are fully occupied in the bulk, where the gap is dominated by $\Delta_Z$. This order of 
the spin-valley splitting is drastically altered at the edge where the confinement potential (\ref{MassTerm}) dominates and where,
according to Eq. (\ref{MassSpec0}), the subbranches ($K,\ua$) and ($K,\da$) are strongly bent upwards, whereas ($K',\ua$) and 
($K',\da$) are bent downwards. As a consequence, the branches ($K',\ua$) and ($K,\da$) cross the Fermi level at the edges and yield
two counter-propagating edge states [Fig. \ref{fig04}(a)]. 
Notice that the chirality of the edge states is determined by the spin orientation, but the counter-propagating states are no longer 
spatially separated. If the two modes are (locally) coupled, e.g. by magnetic scatterers, one is therefore
confronted with a very particular {\textit{dissipative}} quantum Hall effect \cite{abanin}.

The situation is quite different if the valley splitting dominates in the bulk, $\Delta_K > \Delta_Z$, in which case the
($K,\ua$) and ($K,\da$) are fully occupied at $\nu=0$ [Fig. \ref{fig04}(b)]. When approaching the edges, the confinement
potential simply adds up to the term $\Delta_K$, which is effectively enhanced at the edge, but does not alter 
the filling of the subbranches. No levels cross the Fermi energy, and the system therefore remains insulating also at the
edges. 

Which of the two scenarios (quantum Hall effect or insulator) describes correctly the physical at $\nu=0$ remains an open question
in today's research, and the answer is probably not universal. Whereas early experiments on exfoliated graphene on a SiO$_2$
substrate hinted at the quantum Hall scenario with a predominant Zeeman effect \cite{zhang2}, more recent experiments on
suspended samples favour the insulator scenario with $\Delta_K>\Delta_Z$ \cite{du}. In contrast to the IQHE at $\nu=\pm 2(2n+1)$
the physical properties at $\nu=0$ therefore seem to depend sensitively on sample quality and the dielectric environment.

\section{Conclusions}

In conclusion, we have reviewed some basic aspects of the IQHE in graphene as compared to the non-relativistic IQHE in
semi-conductor heterostructures with a parabolic band. The effect is universal with respect to its essential
ingredients, which are, in the context of diffusive samples, (1) the energy quantisation into highly degenerate 
but well separated Landau levels, (2) (semi-classical) localisation due to impurities in the bulk, and (3) chiral current-carrying
edges that contribute $g e^2/h$ per Landau level to the conductance. Graphene reveals, however, significant differences with
respect to conventional 2D electron systems. Most saliently, Landau quantisation yields energy levels that disperse differently
($\propto \lambda\sqrt{Bn}$) in graphene as well as a particular zero-energy level that is only half-filled at the 
charge neutrality point ($\nu=0$). The physical properties of graphene at this particular point depend on the hierarchy 
of the spin-valley degeneracy lifting -- if the Zeeman effect dominates the $n=0$ level splitting in the bulk, the formation
of two counter-propagating edge states yields a particular dissipative quantum Hall effect, whereas a predominant valley splitting does
not provide current-carrying edge states such that the system is insulating.

In this short review, we have hardly discussed the role of Coulomb interactions. Apart from a generalised quantum Hall ferromagnet,
these interactions give rise to the FQHE that has recently been observed in the two-terminal \cite{du,bolotin}
and four-terminal configurations \cite{kim1,kim2}. These observations seem to corroborate a particular four-component form of the
FQHE, which had previously been studied theoretically \cite{toke,GR}, on the basis of the approximate SU(4) spin-valley symmetry of the 
Coulomb interaction \cite{goerbigRev}.

\section*{Acknowledgments}

The author acknowledges fruitful discussions with J.-N. Fuchs, P. Lederer, G. Montambaux, and N. Regnault.



\end{document}